\documentclass[aps,prd,reprint,twocolumn,superscriptaddress,nofootinbib]{revtex4-2}
\usepackage{xcolor}
\usepackage[sort&compress]{natbib}
\usepackage{amsmath,amssymb,graphicx,bm,bbm,slashed,subdepth,amsfonts,physics}
\usepackage{xr-hyper}
\usepackage{subcaption}
\usepackage[colorlinks=true
,urlcolor=blue
,anchorcolor=blue
,citecolor=blue
,filecolor=blue
,linkcolor=red
,menucolor=blue
,linktocpage=true
,pdfproducer=medialab
,pdfa=true
]{hyperref}
\usepackage{caption}
\usepackage{cleveref}
\usepackage{enumerate}
\usepackage{epsfig}
\usepackage{setspace}
\usepackage{booktabs, tabularx}
\usepackage{units}
\usepackage{placeins}
\usepackage{multirow}
\usepackage{mathtools}
\usepackage{xcolor}
\usepackage[normalem]{ulem}

\graphicspath{{figures/}}
\usepackage[english]{babel}
\makeatletter
\addto\captionsenglish{\def\language@en{}}
\let\l@en\l@english
\makeatother

\newcommand{\nn}{\nonumber}

\begin{document}

\title{Signatures of Ultralight Dark Matter in Space-Based Laser Interferometers}

\author{Tingyuan Jiang}
\affiliation{University of Chinese Academy of Sciences (UCAS), Beijing 100049, China}
\affiliation{International Center for Theoretical Physics Asia-Pacific, Beijing, China}
\author{Yong Tang}
\affiliation{University of Chinese Academy of Sciences (UCAS), Beijing 100049, China}
\affiliation{International Center for Theoretical Physics Asia-Pacific, Beijing, China}
\affiliation{School of Fundamental Physics and Mathematical Sciences, \\Hangzhou Institute for Advanced Study, UCAS, Hangzhou 310024, China}

\begin{abstract}
Ultralight dark matter (ULDM) coupled to the Standard Model may effectively induce coherent oscillations of fundamental constants and thereby generate narrow-band signals in precision interferometric experiments. Here we present a systematic study of how these oscillations leave distinctive imprints on space-based laser interferometers, including LISA and Taiji. Starting from the one-way inter-spacecraft link observables, we analyze several instrument-level effects induced by ULDM, including composition-dependent acceleration of test masses, laser-frequency variations associated with cavity-length modulation, refractive-index effects, and clock-related contributions. We then propagate these signals through the standard data processing chain, including time-delay interferometry and clock-noise elimination. We show that the observability of an ULDM-induced effect is determined by the structure of its single-link response. In particular, the ULDM-driven variation in laser frequency appears in the raw link observable with the same form as laser phase noise. As a consequence, it is strongly suppressed in the final interferometry channels. In contrast, signals that possess an explicit directional pattern—for example, ULDM-induced oscillations of the test masses—are not eliminated by this procedure. We further construct a local observable that isolates the differential motion between the test mass and the optical bench, and derive its sensitivity to both the dilaton--gluon coupling $d_g$ and the dilaton--electron coupling $d_e$ for LISA, Taiji, and BBO. We find that the local observable yields sensitivities comparable to the standard Michelson interferometer for $d_g$, but better than Michelson channel by three orders of magnitude for $d_e$. 
\end{abstract}

\date{\today}

\maketitle

\section{Introduction}

Astrophysical and cosmological observations supporting the existence of dark matter (DM) have steadily accumulated since the last century. Nevertheless, the fundamental physical nature of DM is still unclear, and numerous microscopic models have been proposed, including several well-motivated scenarios in which DM is composed of an ultralight bosonic field. In the regime of high occupation number, such a field effectively acts as a coherent classical background and may be described as an oscillating condensate on both laboratory and astrophysical scales, offering a phenomenologically rich set of observational signatures~\cite{hui_ultralight_2017, marsh_axion_2016, Ferreira:2020fam, Chavanis:2025qcg}. If ultralight dark matter (ULDM) interacts with Standard Model (SM) particles, it may effectively generate oscillatory variations in fundamental constants, such as the fine-structure constant, fermion masses, and the hadronization scale~\cite{flambaum_variation_2007, damour_equivalence_2010}. Such signals are characterized by narrow bandwidths and long coherence times, making them accessible to precision experiments.

This possibility has motivated an intensive experimental program searching for oscillating fundamental constants across a wide range of platforms. Atomic and optical clocks provide some of the most sensitive probes: comparisons between different atomic species over extended periods allow the direct detection of modulated shifts in atomic transition frequencies, yielding stringent limits on ULDM couplings to photons, electrons, and light quarks~\cite{derevianko_hunting_2014, arvanitaki_searching_2015, doi:10.1126/sciadv.aau4869, PhysRevLett.130.253001, kennedy_atomic_2020}. Similarly, laser and maser interferometers have been proposed as probes of scalar and pseudoscalar ULDM, because their observables---optical phase or fractional-frequency comparisons over long baselines---are naturally sensitive to oscillations in refractive indices, material dimensions, and test-mass (TM) positions~\cite{stadnik_searching_2015, aoki_detecting_2016, arvanitaki_search_2018, grote_novel_2019, nagano_axion_2019}. Ground-based gravitational-wave (GW) detectors such as LIGO, Virgo, and GEO600 have already been used to set direct limits on scalar and vector ULDM couplings through their laser-frequency and TM-displacement responses~\cite{pierce_searching_2018, vermeulen_direct_2021, morisaki_improved_2021, miller_probing_2021, hall_advanced_2022, michimura_ultralight_2020, guo_darkphoton_2019}.

Space-based GW detectors---such as LISA~\cite{LISA}, Taiji~\cite{Taiji, ruan_taiji_2020}, and TianQin~\cite{TianQin}---extend the search for ULDM into the millihertz frequency band, where they operate with million-kilometer baselines and achieve sensitivity to coupling strengths inaccessible to ground-based experiments. A growing body of literature has investigated the prospects for ULDM detection with these instruments, including: sensitivity forecasts for scalar DM coupled to tess masses (TMs) through dilaton charges~\cite{morisaki_detectability_2019, yu_sensitivity_2023, yao_probing_2024}, gravitational signatures of ULDM through direct metric perturbations~\cite{kim_gravitational_2023, yu_detecting_2024, zhang_probing_2025}, axion-photon coupling effects in the form of polarization rotation and propagation-phase modulation~\cite{yao_prospects_2025, gue_probing_2025, yao_axion-like_2025, liu_detectability_2026}, and the first experimental bounds from LISA Pathfinder on dark photon DM~\cite{frerick_riding_2024}. The LISA Pathfinder mission has also placed constraints on ultralight scalar fields through its precision accelerometry~\cite{miller_lisa_pathfinder_2023}.

Despite these extensive activities, an important but more subtle issue has so far attracted relatively limited attention in discussions of ULDM searches using space-based detectors, namely, whether those ULDM induced effects on lasers, clocks and other instruments are also observables. For instance, in ground-based interferometers such as LIGO, the dominant laser phase noise is suppressed by the common-mode rejection of equal-arm Michelson interferometry. For space-based detectors, however, the arm lengths are unequal and continuously varying, so that simple interferometry does not cancel the laser phase noise. Instead, the raw one-way inter-spacecraft phase measurements must be processed through time-delay interferometry (TDI)~\cite{tinto_time-delay_2020, armstrong_time-delay_1999, estabrook_time-delay_2000, Wang:2020pkk}, which synthesizes equal-path combinations by appropriately delaying and differencing the link data. 
Laser phase noise, which can exceed the GW signal by more than seven orders of magnitude, is thereby reduced to below the secondary noise floor. 
Additionally, the clock noise from onboard ultra-stable oscillators (USOs) is removed via dedicated sideband modulation and subtraction~\cite{otto_tdi_2012, tinto_time-delay_2018, hartwig_clock-jitter_2021}. Consequently, whether an ULDM-induced perturbation is ultimately observable is not determined solely by the existence of a signal at the single-link level; the decisive question is whether the signal structure survives the TDI and clock-noise elimination pipeline that defines the final science observables. 

This question is especially acute for ULDM signals arising from oscillations of fundamental constants. At the single-link level, such oscillations can, in principle, modulate the laser frequency via cavity-length variations and refractive-index changes~\cite{grote_novel_2019}, alter the timing reference provided by the USO~\cite{campbell_searching_2021}, and exert composition-dependent forces on TMs~\cite{morisaki_detectability_2019}. However, different physical mechanisms produce qualitatively different operator structures in the one-way link observables. Some contributions carry directional or propagation-dependent information, whereas others enter the raw data in a form structurally indistinguishable from laser phase noise. Establishing this distinction is essential for identifying which ULDM couplings can actually be probed by TDI-based interferometers and for avoiding overly optimistic projections.

In this work, we investigate how ULDM-induced oscillations of physical constants enter the single-link observables of space-based gravitational-wave detectors and how these contributions propagate through the standard data-processing and interferometric combinations. We consider several instrument-level effects induced by ULDM, including composition-dependent acceleration of the TMs, variations of cavity-stabilized laser frequencies, refractive-index effects in optical components, and clock-related contributions. We show that the detectability of an ULDM signal is controlled by the structure of its single-link response. In particular, the signal associated with laser-frequency variation generated by oscillating fundamental constants is structurally degenerate with laser phase noise at the single-link level and is therefore strongly suppressed or canceled by TDI. By contrast, signals with explicit directional structure---such as ULDM-induced oscillations of the test masses, as well as other propagation-related contributions---are not removed and can remain observable. We further propose a local observable associated with the differential motion between the TM and the optical bench (OB), illustrate its sensitivities to the dilaton--gluon coupling $d_g$ and the dilaton--electron coupling $d_e$, and compare with the standard Michelson interferometer. We find that the local observable is comparable to the standard Michelson channel for $d_g$, but much better for $d_e$ by three orders of magnitude. 
Our work provide a more complete and systematic framework for analyzing the physical effects of ULDM on laser interferometers in space. 

The paper is organized as follows. In Sec.~\ref{sec:theory}, we introduce the ULDM model and describe the properties of the local DM field. In Sec.~\ref{sec:SGWD}, we introduce the measurement scheme of space-based GW detectors, including the single-link observables, noise sources, and the TDI and clock-noise elimination framework. In Sec.~\ref{sec:varyingobject}, we analyze how oscillations of fundamental constants modify specific instruments and generate different classes of single-link signals. In Sec.~\ref{sec:signal and cancel}, we compute the propagation of each signal through TDI and clock-noise calibration, and present the sensitivity of the local observable $\xi$. In Sec.~\ref{sec:analysis}, we formulate a general structural criterion that governs whether an ULDM-induced single-link contribution survives into the final observables. We conclude in Sec.~\ref{sec:conclusion}. Throughout this work, we use natural units with $\hbar=c=1$.

\section{Theoretical Framework}\label{sec:theory}

We start from the Lagrangian for the interaction between a scalar ULDM $\Phi(x)$ and SM~\cite{damour_equivalence_2010}
\begin{equation}\label{eq:model}
    \begin{aligned}
    \mathcal{L}_{\mathrm{int}}&=\varphi\biggl[\frac{d_{e}}{4e^2}(F_{\mu\nu})^{2}-\frac{d_{g}\beta_{3}}{2g_{3}}(G_{\mu\nu})^{2}\\&-\sum_{i=e,u,d}(d_{m_{i}}+\gamma_{m_{i}}d_{g})m_{i}\bar{\psi}_{i}\psi_{i}\biggr],
    \end{aligned}
\end{equation}
where $\varphi=\kappa \Phi$ is a rescaled dimensionless field, $\kappa\equiv\sqrt{4\pi G}$ is the inverse of the Planck mass, $F_{\mu\nu}$, $G_{\mu\nu}$ and $\psi_i$ are the electromagnetic Faraday tensor, gluon field strength tensor and fermion spinors, $d_{e},~d_{g},~d_{m_{i}}$ are dimensionless coupling constants, $g_3$ the gauge coupling of strong interaction (QCD), $\beta_3$ is the $\beta$ function for the running of $g_3$, $\gamma_{m_{i}}$ is the anomalous dimensions of fermion. 

In the presence of ULDM background, the above interactions induce
oscillatory shifts of the SM parameters. In particular, the
fine-structure constant, the fermion masses, and the QCD confinement scale
become functions of the dimensionless scalar field $\varphi$,
\begin{align}
    \alpha(\varphi)&=\alpha\left(1+d_e \varphi\right),\label{eq:aphi}\\
    m_i(\varphi)&=m_i\left(1+d_{m_i} \varphi\right),~\text{for}~ i=e,u,d, \label{eq:miphi}\\
    \Lambda_{\text{QCD}}(\varphi)&=\Lambda_{\text{QCD}}\left(1+d_{g} \varphi\right).\label{eq:lbdphi}
\end{align}
We use renormalization-group-invariant fermion masses, so the effect of the mass anomalous dimension $\gamma_m$ is absorbed into the definition of $m_i$ and does not appear as an independent contribution to the variations above. A detailed derivation of this parametrization can be found in Ref.~\cite{damour_equivalence_2010}.

The local ULDM field $\Phi(t,\boldsymbol{x})$ may be represented as a sum of monochromatic plane waves spanning a range of velocities
\begin{align}
    \Phi(t,\boldsymbol{x})=\sum_{\boldsymbol{v}}\frac{\sqrt{2\rho/N}}{m}\exp[i(\omega t-\boldsymbol{k}\cdot \boldsymbol{x}+\theta_v)],
        \label{eq:sum_phi}
\end{align}
 where $m$ is the mass of ULDM particles, $\rho\approx0.3~\mathrm{GeV/cm^3}$ is the local energy density of DM, $\theta_{\boldsymbol{v}} $ are random phases drawn independently from $[0,2\pi)$ with uniform distribution, and $N$ denotes the total number of ULDM particles in the vicinity. In the non-relativistic limit the field varies slowly, which means $\omega\simeq m(1+\boldsymbol{v}^2/2)$, $\boldsymbol{k}\simeq m\boldsymbol{v}$, the fast oscillations can be factored. Therefore the behavior of the ULDM can be described by
 \begin{align}
     \Phi(\boldsymbol{x},t)=\Phi_0(\boldsymbol{x},t)\exp{i\left[mt+\theta_0(\boldsymbol{x},t)\right]},\label{eq:phixt}
 \end{align}
where the amplitude $\Phi_0(\boldsymbol{x}, t)$ and phase $\theta_0(\boldsymbol{x}, t)$ have the property of varying stochastically on the scales of coherence length $\lambda_c$ and coherence time $\tau_c$
\begin{align}
    \tau_{c}&=\frac{2\pi}{m\sigma^2}\approx4.13\times10^8\mathrm{s}\left(\frac{10^{-17}\mathrm{eV}}{m}\right),\\
    \lambda_{c}&=\frac{2\pi}{m\sigma}\approx1.24\times10^{11}\mathrm{km}\left(\frac{10^{-17}\mathrm{eV}}{m}\right),
\end{align}
where $\sigma\simeq10^{-3}$ is the DM velocity dispersion. Within one coherence patch, the stochastic amplitude and phase vary only on
the scales $\lambda_c$ and $\tau_c$. In the mass range relevant for space-based gravitational-wave detectors (SGWDs), both the detector size and the observation time are smaller than the coherence length and coherence time, respectively. We can therefore neglect the slow stochastic variation of the envelope. However, we keep the deterministic plane-wave phase $-\boldsymbol{k}\cdot \boldsymbol{x}$, since it is responsible for the spatial gradient of the ULDM field. The local field can then be approximated by a single coherent plane-wave component,
\begin{align}\label{eq:phicos}
    \Phi(t)=\Phi_0\exp[i\left(mt-\boldsymbol{k}\cdot \boldsymbol{x}+\theta_0\right)],
\end{align}
where $\Phi_0=\sqrt{2\rho}/m$. This expression should be understood as the complex representation of the real ULDM field, with the physical field obtained by taking the real part.

Equation~\eqref{eq:phicos} implies that the local ULDM field induces periodic modulations of fundamental constants. These oscillations can, in principle, generate observable signatures in a variety of precision experiments, including SGWDs~\cite{grote_novel_2019}. In such instruments, DM field may impart additional acceleration to material bodies, alter material dimensions or refractive indices of optical components, and thereby shift derived instrumental quantities. We shall return to these detector-specific couplings and their parametric impacts in Sec.~\ref{sec:varyingobject} after a brief introduction of SGWDs below.

\section{Measurements of SGWDs}\label{sec:SGWD}

\begin{figure}[t]
    \centering
    \includegraphics[width=0.8\linewidth]{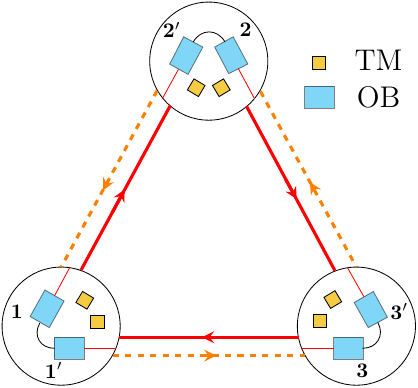}
    \caption{Schematic SGWD configuration. The three large circles denote the spacecrafts, labeled by $1$, $2$, and $3$, each sending and receiving laser beams from other two. Each spacecraft contains two optical benches, shown as blue rectangles and labeled by unprimed and primed indices $(1,1')$, $(2,2')$, and $(3,3')$. The yellow blocks denote the test masses (TM) associated with the corresponding optical benches. 
    The thin red lines inside each spacecraft indicate the local optical links used to compare the two benches and the associated test-mass readouts.}
    \label{fig:SGWD_path}
\end{figure}

To provide a clearer explanation of how DM oscillations generate signals, we briefly introduce the SGWD measurement scheme in this section.

SGWDs such as LISA and Taiji can be viewed as very large Michelson-like interferometers, with the arms formed by laser links between spacecraft, see Fig.~\ref{fig:SGWD_path}. LISA uses Nd:YAG laser systems at a wavelength of $\lambda \simeq 1064~\mathrm{nm}$, corresponding to an optical carrier frequency $f_0=2\pi \nu_0=c/\lambda \simeq 3\times10^{14}$~Hz. Each spacecraft carries two local lasers that are frequency-stabilized, for example to an optical cavity, and the inter-spacecraft links are formed by exchanging and heterodyning these lasers. Three spacecrafts fly in a triangular heliocentric orbit, with arm lengths of order $10^6$ km. Each spacecraft also carries two free-falling TMs, which act as nearly perfect inertial references, while the spacecraft use drag-free control to follow them. A passing GW changes the light-travel time along the arms, which is recorded as a change in the measured optical phase.

\begin{figure}[t]
    \centering
    \includegraphics[width=1.0\linewidth]{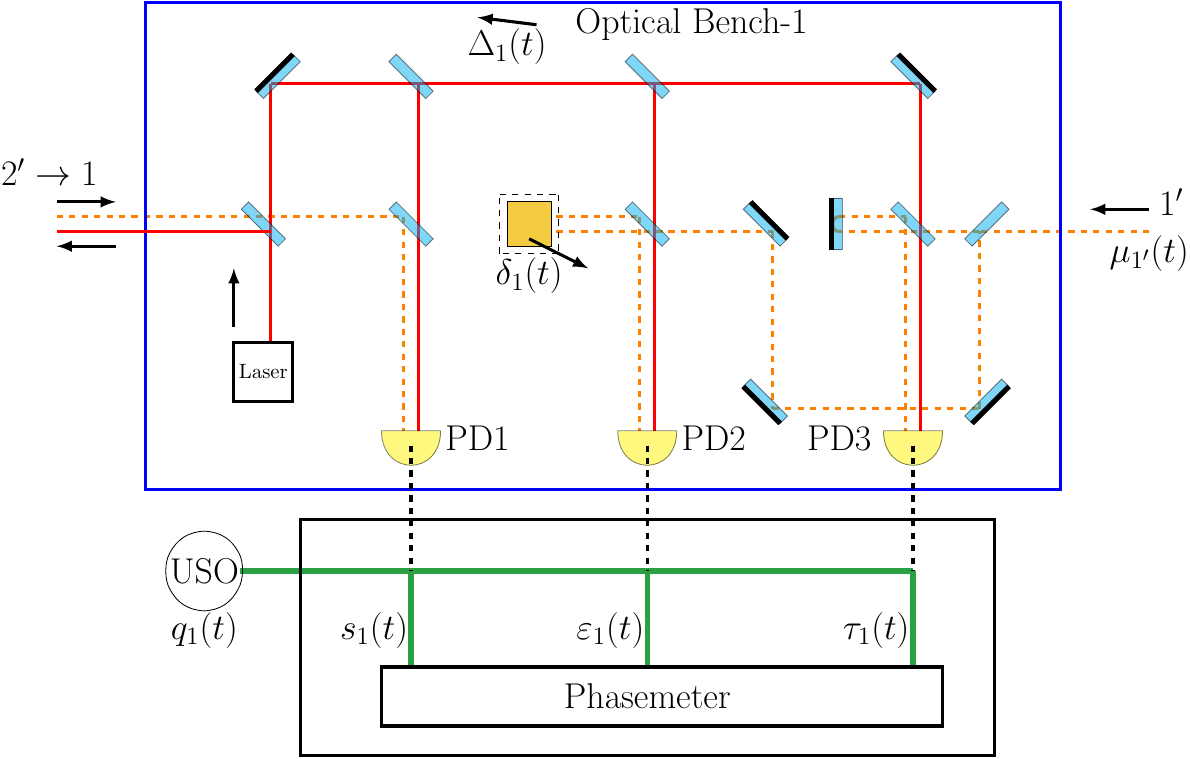}
    \caption{Schematic layout of one optical bench. $s_1(t)$, $\varepsilon_1(t)$, and $\tau_1(t)$ denote the science signal, the TM interferometer signal, and the reference output, respectively.}
    \label{fig:InterferometerOB}
\end{figure}

The basic measurement is a one-way phase signal on each link, or equivalently fractional frequency. Light sent from spacecraft $s$ is received at spacecraft $r$ and interfered with the local laser on $r$. 
The optical path is shown in Fig.~\ref{fig:SGWD_path}. The resulting beat note is tracked by a phasemeter. Because the spacecrafts move relative to each other, the received frequency is Doppler shifted, and the beat note typically lies at radio frequencies, often in the MHz range. Fig.~\ref{fig:InterferometerOB} shows a sketch of one OB. Three measurements, $s$, $\varepsilon$, and $\tau$, are performed on each OB. Their phase readouts contain the gravitational-wave response, laser phase noise, clock noise, and other measurements and device noises. For a more complete signal model, see Ref.~\cite{otto_tdi_2012}.

In addition to the gravitational-wave signal, these measurements contain several noise contributions, which can be grouped into four standard classes:
(1) TM acceleration noise (proof-mass noise), which dominates at low frequencies.
(2) Optical readout noise (including shot noise and other optical-path/measurement noises), which dominates at high frequencies.
(3) Laser frequency (phase) noise, which is exceptionally large in raw data; unlike ground-based Michelson interferometers, the arms in LISA/Taiji are unequal and time-dependent, so this noise does not cancel by a simple interferometry.
(4) Clock noise, since an onboard ultra-stable oscillator (USO) provides the timing and frequency reference for the phasemeter; fluctuations of this reference can couple into the measured phases.

To demonstrate various contributions in detail, we show the expressions for the measurements in OB(1)~\cite{tinto_time-delay_2018},
\begin{align}
    s_1=&\left[h_{1}+\mathcal{D}_{12}p_{2^{\prime}}-p_1-2\pi\nu_{2^{\prime}}\left(\mathbf{n}_{12}\cdot\mathcal{D}_{12}\mathbf{\Delta}_{2^{\prime}}\right.\right.\nn\\
    &\left.\left.+\mathbf{n}_{21}\cdot\mathbf{\Delta}_1\right)+N_1^\mathrm{opt}\right]-a_1q_1+N_1^{s},\label{eq:noiseonOBs}\\
    \varepsilon_1=&\left[p_{1^{\prime}}-p_1+4\pi\nu_{1^{\prime}}\left(\mathbf{n}_{21}\cdot\boldsymbol{\delta}_1-\mathbf{n}_{21}\cdot\boldsymbol{\Delta}_1\right)+\mu_{1^{\prime}}\right]\nn\\
    &-b_1q_1+N_1^\varepsilon,\label{eq:noiseonOBe}\\
    \tau_1=&\left[p_{1^{\prime}}-p_1+\mu_{1^{\prime}}\right]-b_1q_1+N_1^\tau.\label{eq:noiseonOBt}
\end{align}
where $\mathcal{D}_{rs}$ is the time-delay operator, defined by $\mathcal{D}_{rs}f(t)\equiv f(t-L_{rs})$; $h_1$ is the signal we are concerned about; $p_i$ and $\nu_i$ denote the laser phase noise and laser frequency on OB($i$), respectively; $\mathbf{n}_{ij}$ is the unit vector pointing from $j$ to $i$; the $\boldsymbol{\Delta}$ and $\boldsymbol{\delta}$ terms represent the OB displacement noise relative to an inertial frame and the TM displacement noise, respectively, as illustrated in Fig.~\ref{fig:InterferometerOB}; the $N^{{s},\varepsilon,\tau}_i$ terms are shot-noise phase fluctuations at the photodetectors and $N^{\mathrm{opt}}_i$ is optical-path noise; $\mu_i$ is fiber noise, assumed to be independent of the propagation direction of the optical beams within the fibers; and the $q$ terms are the clock phase noises due to the three USOs. The corresponding coefficients are
\begin{align}
    a_1&=\frac{\nu_{2^\prime}(1-\dot{L}_{12})-\nu_1}{f_{\mathrm{USO}1}}\label{eq:noisea},\\
    b_1&=\frac{\nu_{1^{\prime}}-\nu_1}{f_{\mathrm{USO}1}},\label{eq:noiseb}
\end{align}
where $f_{\mathrm{USO}i}$ are the USO pilot-tone frequencies \cite{otto_tdi_2012}.

We can eliminate OB noise by combining these  measurements:
\begin{align}\label{eq:eta1}
    &\eta_{12}\equiv s_{1}-\frac{\nu_{2^{\prime}}}{\nu_{1^{\prime}}}\frac{(\epsilon_{1}-\tau_{1})}{2}\nn\\&-\frac{\nu_{2^{\prime}}}{\nu_{2}}\mathcal{D}_{12}\frac{(\epsilon_{2^{\prime}}-\tau_{2^{\prime}})}{2}-\mathcal{D}_{12}\frac{\tau_2-\tau_{2^{\prime}}}{2}.
\end{align}
Along with the other five $\eta_{ij}$ data streams, these constitute the preprocessed single-link measurements.

Optical Metrology System (OMS) noise, composed of optical-path noise $N^{\mathrm{opt}}$ and shot noise $N^{{s},\varepsilon,\tau}_i$, together with TM acceleration noise, can be modeled following~\cite{babak_lisa_2021}.
Laser phase noise and clock noise, however, require dedicated cancellation methods.
To deal with laser phase noise, TDI~\cite{tinto_time-delay_2020} is used in LISA and Taiji (see Appendix~\ref{appendix:X}) to combine the one-way link measurements after shifting them by the measured light-travel times, so that the laser phase noise is canceled to high accuracy. For clock noise, modulating the carrier laser with USO-referenced microwave sidebands and measuring both the carrier-carrier and sideband-sideband beat notes provides an independent estimate of the inter-spacecraft clock phase difference, which can then be subtracted from the science measurement through standard USO calibration \cite{otto_tdi_2012,tinto_time-delay_2018,hartwig_clock-jitter_2021}.

\section{Individual Effects}\label{sec:varyingobject}
In this section, we analyze the influence of ULDM on each instrument, treating lasers, clocks, and TMs individually. As will be demonstrated in the following section, some of these effects are highly suppressed in the final interferometry channels.

\subsection{Effect on Laser frequency}

Oscillations of the fundamental constants induced by ULDM affect the detector readout through two closely related channels. First, they modify the characteristic material length scale, which changes the size of solid components such as the spacer cavity and optical elements. Second, they alter the electromagnetic response of dielectric materials, thereby changing the refractive index and hence the optical path inside the OB. These effects together lead to an effective fractional-frequency variation that enters the one-way inter-spacecraft measurement.

To first order, and ignoring small relativistic corrections, the length of a solid is proportional to the Bohr radius $a_\text{B}$, the characteristic atomic length scale \cite{PhysRevLett.122.160801,Griffiths_Schroeter_2018}. Thus, for solids such as ultra-low-expansion (ULE) spacer cavities and oven-controlled crystal oscillators,
\begin{align}
    L_{\mathbf{c}}\propto a_\text{B}\propto\alpha^{-1}m_{e}^{-1},
    \label{eq:L_a_m}
\end{align}
where $L_{\mathbf{c}}$ denotes the length of the spacer cavity. If $\alpha$ and $m_e$ vary with time, the material expands or contracts accordingly. Since we are focusing on ULDM oscillation frequency $f_c=m/2\pi$ which is much smaller than a typical fundamental vibrational frequency of the solid, $f_0\sim10^4\,\mathrm{Hz}$, the response can be treated adiabatically \cite{grote_novel_2019}. Substituting Eqs.~\eqref{eq:aphi} and \eqref{eq:miphi} into Eq.~\eqref{eq:L_a_m}, we obtain
\begin{align}
    \frac{\delta L_{\mathbf{c}}}{L_{\mathbf{c}}}=-\frac{\delta \alpha}{\alpha}
        -\frac{\delta m_e}{m_e}
    =
    -\kappa\left(d_e+d_{m_e}\right)\Phi.
    \label{eq:Lphi}
\end{align}

The laser frequency is usually stabilized by the Pound--Drever--Hall (PDH) technique \cite{drever_laser_1983}, which locks the laser to the resonance frequency of the ULE reference cavity. Therefore, a change in the cavity length directly induces a corresponding fractional shift of the laser frequency,
\begin{align}
    \frac{\delta \nu_{\mathbf{laser}}}{\nu_{\mathbf{laser}}}
    =
    -\frac{\delta L_{\mathbf{c}}}{L_{\mathbf{c}}}.
    \label{eq:nuL}
\end{align}

Besides changing the material length scale, oscillations of $\alpha$ and $m_e$ also modify the optical properties of the components on the OB. In particular, the beam splitters and phase shifters used in the modulation and routing of the laser field acquire a time-dependent refractive index. To estimate this effect, we adopt a simple Lorentz model with a single electronic mode of frequency $\nu_0$ and assume that the laser frequency is much larger than all phonon-mode frequencies of the beam splitter. The relative permittivity is then \cite{grote_novel_2019}
\begin{align}
    \varepsilon_r
    \approx
    1+\frac{\xi N\alpha}{4\pi^2 m_e}\frac{1}{\nu_0^2-\nu_{\mathbf{laser}}^2},
\end{align}
where $N$ is the number density of atoms in the dielectric material and $\xi$ is a numerical constant independent of the fundamental constants. Neglecting magnetic permeability, the refractive index is
\begin{align}
    n\approx\sqrt{\varepsilon_r}.
\end{align}
The combination $\xi N\alpha/(4\pi^2 m_e \nu_0^2)$ is independent of the fundamental constants because $N\propto a_\text{B}^{-3}$ and $\nu_0\propto m_e\alpha^2$. Therefore,
\begin{align}
    \frac{\delta n}{n}
    \approx
    \frac{\nu_{\mathbf{laser}}}{n}\frac{\partial n}{\partial \nu_{\mathbf{laser}}}
    \left(
        \frac{\delta \nu_{\mathbf{laser}}}{\nu_{\mathbf{laser}}}
        -2\frac{\delta \alpha}{\alpha}
        -\frac{\delta m_e}{m_e}
    \right).\label{eq:delta_n1}
\end{align}
For a laser wavelength $\lambda=1064\,\mathrm{nm}$ and a silica beam splitter, one has approximately $n\approx1.5$ and
\[
\frac{\nu_{\mathbf{laser}}}{n}\frac{\partial n}{\partial \nu_{\mathbf{laser}}}\approx 5\times10^{-3}.
\]
Substituting Eq.~\eqref{eq:nuL} into Eq.~\eqref{eq:delta_n1}, we can get
\begin{align}
    \frac{\delta n}{n}
    \approx
    -5\times10^{-3}\frac{\delta\alpha}{\alpha}.\label{eq:delta_n2}
\end{align}

The same oscillation of fundamental constants also changes the geometric size of the optical element itself. Hence the optical path inside the beam splitter,
\[
L_{\mathrm{op}}=nl,
\]
with $l$ the beam-splitter thickness, varies according to
\begin{align}
    \frac{\delta L_{\mathrm{op}}}{L_{\mathrm{op}}}
    =
    \frac{\delta n}{n}
    +
    \frac{\delta l}{l}.
\end{align}
Therefore, the ULDM-induced modulation of physical constants affects the measured phase in two ways: through the cavity-stabilized laser frequency and through the optical path of the optical elements on the bench. Combining these two effects, the effective fractional-frequency variation entering the measurement can be written as
\begin{align}
    \frac{\delta \nu}{\nu}
    =
    -\frac{\delta L_{\mathbf{c}}}{L_{\mathbf{c}}}
    -\frac{\delta L_{\mathrm{op}}}{L_{\mathrm{op}}}
    \approx
    2.005\frac{\delta \alpha}{\alpha}
    +2\frac{\delta m_e}{m_e}.
    \label{eq:nuam}
\end{align}
This effective variation is already present at the optical-bench level and therefore appears directly in the single-link observable.

The corresponding ULDM-induced signal on one link from spacecraft $s$ to spacecraft $r$ is then
\begin{align}
    \eta_{rs}(t)
    &=
    \frac{\delta\nu_r(t)-\mathcal{D}_{rs}\delta\nu_s(t)}{\nu}
    \label{eq:etanu}\\
    &\approx
    2\kappa\left(d_e+d_{m_e}\right)
    \left[
        \Phi(t,x_r)-\mathcal{D}_{rs}\Phi(t,x_s)
    \right].
    \label{eq:etaPhi}
\end{align}


\subsection{Clock jitter}

Apart from the laser cavity, the USO is also an important component in SGWDs. It is therefore natural to expect that the reference frequency supplied by the USO is also modulated by ULDM, which may consequently introduce an additional effect into the signal measurement. We consider the widely adopted quartz oscillator as the USO reference; unlike a ULE cavity, it produces a stable frequency through a bulk acoustic wave. The dependence of the quartz-oscillator mode on the fundamental constants is given by \cite{campbell_searching_2021}
\begin{align}
    &f_{\text{USO}}\propto m_e\alpha^2\sqrt{\frac{m_e}{\Lambda_{\mathrm{QCD}}}},\\
    &\frac{\delta f_{\text{USO}}}{f_{\text{USO}}}\simeq \kappa\left(2d_e+\frac{3}{2}d_{m_e}-\frac{1}{2}d_g\right)\Phi.
\end{align}

ULDM can affect the clock through clock jitter. Let $\tau$ denote the proper time along the spacecraft worldline and $\hat{\tau}_i$ the time kept by the USO on OB(i). We parameterize the USO timing error by
\begin{align}
    \hat{\tau}_i=\tau+\tilde{q}_i\left(\tau\right),
\end{align}
where $\tilde{q}_i$ is the clock jitter. Standard treatments assume $\tilde{q}_i$ is a stochastic process, usually modelled as zero-mean Gaussian noise; here we use frequency variation to represent that:
\begin{align}
    \tilde{q}_i\left(\tau\right)=\int^\tau d\tau^{\prime}\frac{\delta f_{\text{USO}}(\tau^{\prime})}{f_{\text{USO}}}.\label{eq:q-fUSO}
\end{align}

The clock noise $q_i$ in Eq.\eqref{eq:noiseonOBs}-\eqref{eq:noiseonOBt} can be expressed as
\begin{align}
    q_i=2\pi f_{\text{USO}}\tilde{q}_i\left(\tau\right).
\end{align}
Therefore, the single-link signal $\eta_{rs}(t)$ defined by Eq.\eqref{eq:eta1}, also receives an ULDM contribution from clock jitter.

\subsection{Acceleration of TMs}

By coupling directly to matter, $\Phi$ can affect the motion of TMs. The action for the TM moving in the ULDM background is linearly given by
\begin{align}
    S=-\int m(\Phi)\sqrt{-\eta_{\mu\nu}dx^\mu dx^\mu}.
\end{align}

Under the nonrelativistic approximation, the equation of motion becomes 
\begin{align}\label{eq:EOM}
    \frac{d^2x}{dt^2}\simeq-\kappa\alpha_{m}(\Phi)\nabla\Phi,\quad\alpha_{m}(\Phi)\equiv\frac{\partial\ln\left[\kappa m(\Phi)\right]}{\partial(\kappa\Phi)},
\end{align}
The form of $\alpha_m(\Phi)$ depends on how ULDM couples to the SM. For the model considered here, Eq.~\eqref{eq:model} gives~\cite{morisaki_detectability_2019}
\begin{align}\label{eq:amdQ}
    \alpha_m(\Phi)\simeq d_g+[(d_{\hat{m}}-d_g)Q_{\hat{m}}+d_eQ_e],
\end{align}
where 
\begin{align}
    d_{\hat{m}}\equiv\frac{d_{m_d}m_d+d_{m_u}m_u}{m_d+m_u},
\end{align}
and $m_{u,d}$ are current quark masses. Here $Q$ denotes the dilaton charge and depends on the material composition through
\begin{align}
Q_{\hat{m}}=F_{A}\biggl[&0.093-\frac{0.036}{A^{1/3}}-0.020\frac{(A-2Z)^{2}}{A^{2}}\nn\\
&-1.4\times10^{-4}\frac{Z(Z-1)}{A^{4/3}}\biggr]\label{eq:Qm},\\
Q_e  = F_A\biggl[&-1.4+8.2\frac{Z}{A}+7.7\frac{Z(Z-1)}{A^{4/3}}\biggr]\times10^{-4}. \label{eq:Qe}
\end{align}
where $Z$ denotes the atomic number, $A$ is the mass number, $F_A=A m_{0}/m_A$, $m_{0}=931~\text{MeV}$ is the atomic mass unit, and $m_A$ is the atomic mass.

An additional acceleration causes nominally free-falling TMs to deviate from geodesic motion. Consequently, TMs hosted on different OBs experience relative accelerations, which generate Doppler shifts on the inter-spacecraft laser links. Since the basic observable in LISA-like detectors is the one-way fractional-frequency fluctuation, and the final science channels are formed by combining these single-link measurements through TDI, the ULDM-induced effect can be treated naturally at the level of the single-link response. In the case of scalar ULDM coupled to the SM, the oscillating field coherently drives test-mass motion through effective-mass modulation and the associated gradient-induced acceleration, that is Eq.~\eqref{eq:EOM}, thereby producing a measurable line-of-sight Doppler signal on each optical link. One may therefore define the ULDM single-link Doppler response in the same way as for gravitational-wave detection,
\begin{align}\label{eq:eta_nrs}
    \eta_{rs}(t)\equiv\frac{\delta\nu}{\nu_0}=-\hat{n}_{rs}\cdot\left(\left.\frac{\dd \boldsymbol{x}}{\dd t}\right|_{t,\boldsymbol{x}_r}-\mathcal{D}_{rs}\left.\frac{\dd \boldsymbol{x}}{\dd t}\right|_{t,\boldsymbol{x}_s}\right),
\end{align}
where $\boldsymbol{x}$ is the TM position and $\hat{n}_{rs}$ is the unit vector pointing from $s$ to $r$. Consequently, ULDM signals of this type can be treated within the same response formalism used for GWs at the level of the link observables, and can then be processed to TDI combinations for data analysis in the full LISA constellation.


\section{Physical Effects after TDI}\label{sec:signal and cancel}
Above we have discussed how various effects are inherited in single-link measurements $\eta_{ij}(t)$. However, these quantities are dominated by the laser phase noises $p_i$ and $p_{i'}$, which are $6-8$ order-of-magnitude larger than the interested signals $h_i$. We need to suppress the laser phase noises by constructing virtual equal-arm interferometers by combining the time-delayed data series $\eta_{ij}(t)$. The details of the constructions are given in Appendix.~\ref{appendix:X}. For our purposes, the Michelson channel $X(t)$ below shall be sufficient for the later discussions.

\subsection{Signal of laser-frequency variation}


We use the standard Michelson channel $X$ for the illustration,
\begin{align}\label{eq:X15Michelson}
    X
    =&
    \left(1-\mathcal{D}_{12}\mathcal{D}_{21}\right)
    \left(\eta_{31}+\mathcal{D}_{31}\eta_{13}\right)\nn\\
    &-
    \left(1-\mathcal{D}_{13}\mathcal{D}_{31}\right)
    \left(\eta_{21}+\mathcal{D}_{21}\eta_{12}\right),
\end{align}
where $\mathcal{D}_{ij}f(t)\equiv f(t-L_{ij}(t))$ is the delay operator. We have calculated that the ULDM-induced contribution is highly suppressed in $X$, even absent in the static case when $L_{ij}(t)$ do not depend on time $t$. 

The reason is that, in the static case the delay operators commute, so the two synthesized Michelson paths sample the same scalar-induced frequency shift with opposite signs. The cancellation is therefore exact, and no residual ULDM signal remains in the $X$ observable.

For time-dependent arm lengths, the cancellation becomes imperfect. Applying two delays in different orders now corresponds to evaluating the ULDM-induced frequency shift at slightly different times. The residual signal is therefore controlled by the non-commutativity of the time-delay operators, or equivalently by the time derivatives of the arm lengths, $\dot L_{ij}$. Hence the surviving contribution is suppressed by one power of $\dot L$. For LISA/Taiji, where $\dot L\sim 10^{-9}$, this gives a strong suppression of the observable effect.

One can further construct a higher-order Michelson observable by arranging the delayed data streams such that the two effective optical paths agree not only for static arms, but also up to first order in the arm-length variation. In that case, the same ULDM-induced contribution cancels through both zeroth and first order in $\dot L$. Therefore, after neglecting terms of order $\dot L^2$ and higher, no physically extractable ULDM signal remains in this higher-order Michelson observable. The explicit expansion of the relevant Michelson combinations is given in Appendix~\ref{appendix:X}.

\subsection{Clock noise reduction}

Clock noise provides another possible channel through which the signal could enter the data. We therefore briefly describe the standard clock-noise elimination procedure and then explain why any ULDM-induced contribution that enters only as clock jitter is removed by the same procedure.

The onboard clock, or ultrastable oscillator (USO), is used as the phase reference for the phasemeter. Its phase fluctuation, denoted by $q_i$, enters the one-way heterodyne measurement because the beat note is measured against a USO-referenced microwave signal. To monitor this clock noise, one modulates the laser with a USO-referenced sideband. The sideband-sideband measurement then carries the remote USO phase noise after propagation, together with the local USO phase noise at reception. Thus the sideband data provide an independent measurement of delayed clock-noise differences.

For example, the sideband measurement on optical bench 1 can be written as
\begin{align}
    s_1^{\mathrm{sb}}
    &=\left[h_1+D_{12}p_{2^{\prime}}-p_1+m_{2^{\prime}}D_{12}q_2-m_1q_1\right.\nn\\
    &\quad\left.-2\pi\nu_{2^{\prime}}\left(\mathbf{n}_{12}\cdot\mathcal{D}_{12}\mathbf{\Delta}_{2^{\prime}}+\mathbf{n}_{21}\cdot\mathbf{\Delta}_1\right)+N_1^{\mathrm{opt.sb}}\right]\nn\\
    &\quad-c_1q_1+N_1^{\mathrm{sb}},\label{eq:noiseonOBssb}
\end{align}
where
\begin{align}
    c_1=\frac{(\nu_{2^\prime}+m_{2^\prime}f_{\mathrm{USO}2})(1-\dot{L}_{12})-(\nu_1+m_1f_{\mathrm{USO}1})}{f_{\mathrm{USO}1}}.\label{eq:noisec}
\end{align}
Here $m_i$ is the integer specifying the modulation frequency in units of the USO frequency. The carrier and sideband measurements have the same propagation structure because both arise from the same heterodyne interference between the received beam and the local laser. The difference is that the sideband also carries the USO-referenced modulation terms, such as $m_{2^\prime}D_{12}q_2-m_1q_1$.

The useful elimination observables are obtained by subtracting the sideband measurement from the corresponding carrier measurement. For example,
\begin{align}
    r_{1^{\prime}}&\equiv\frac{{s_{1^{\prime}}}-s_{1^{\prime}}^{\mathrm{sb}}}{m_{3}f_{\mathrm{USO}3}}\simeq\frac{q_1}{f_{\mathrm{USO}1}}-\frac{\mathcal{D}_{13}q_3}{f_{\mathrm{USO}3}},\\
    r_{1}&\equiv\frac{{s_{1}}-s_{1}^{\mathrm{sb}}}{m_{2^{\prime}}f_{\mathrm{USO}2}}\simeq\frac{q_1}{f_{\mathrm{USO}1}}-\frac{\mathcal{D}_{12}q_2}{f_{\mathrm{USO}2}} .
\end{align}
These quantities measure differences between local and delayed remote clock phase noises. Therefore, delayed linear combinations of the $r_i$ can be used to reconstruct and subtract the clock-noise terms appearing in the TDI observable.

For the Michelson channel $X$, the clock-noise-calibrated combination~\cite{hartwig_clock-jitter_2021} is
\begin{widetext}
\begin{align}
    X^{\mathrm{c}}
    &\equiv X-\left[\mathcal{D}_{12}\mathcal{D}_{21}\mathcal{D}_{13}\mathcal{D}_{31}-I\right]\biggl\{b_{1^{\prime}}\frac{f_{\mathrm{USO}1}}{2}\bigl[(I-\mathcal{D}_{12}\mathcal{D}_{21})(r_{1^{\prime}}+\mathcal{D}_{13}r_{3})\nn\\
    &\quad+(I-\mathcal{D}_{13}\mathcal{D}_{31})(r_{1}+\mathcal{D}_{12}r_{2^{\prime}})\bigr]+a_{1}f_{\mathrm{USO}1}(r_{1^{\prime}}+\mathcal{D}_{13}r_{3})-a_{1^{\prime}}f_{\mathrm{USO}1}(r_{1}+\mathcal{D}_{12}r_{2^{\prime}})\nn\\
    &\quad+a_{2^{\prime}}f_{\mathrm{USO}2}\left[r_{1^{\prime}}-(I-\mathcal{D}_{13}\mathcal{D}_{31})r_{1}+\mathcal{D}_{13}r_{3}\right]-a_{3}f_{\mathrm{USO}3}\left[r_{1}-(I-\mathcal{D}_{12}\mathcal{D}_{21})r_{1^{\prime}}+\mathcal{D}_{12}r_{2^{\prime}}\right]\biggr\}.
    \label{eq:X20clockcal}
\end{align}
\end{widetext}
The essential point is that this subtraction uses directly measured clock-noise differences. It does not require a statistical model for the clock noise spectrum; it only assumes that the clock fluctuation enters the carrier and sideband measurements through the same linear readout chain. Therefore, any ULDM-induced contribution that appears only as an additional clock phase fluctuation is removed together with the ordinary USO noise by the same elimination. Within this approximation, clock jitter does not provide an independent residual ULDM signal in the calibrated $X$-channel observable.

\subsection{Sensitivity of TMs' Acceleration}
The sensitivity to ULDM of SGWDs has been systematically investigated in~\cite{yu_sensitivity_2023, yao_probing_2024, Xu:2025rfv, Gue:2025iab, Yao:2025vgy}, based on different interferometry channels constructed from data in three spacecrafts. Here instead we highlight another local quantity in a single spacecraft that can be used to probe ULDM. As we shall see, in some cases this quantity could be more sensitive than the usual interferometers. 

Note that the additional acceleration in Eq.~\eqref{eq:amdQ} depends on the composition of materials. 
Then the relative motion between a TM and its local OB would be an additional observable. Taking a closer look at the TM interferometric signal in Eq.~\eqref{eq:noiseonOBe} and the reference output in Eq.~\eqref{eq:noiseonOBt}, one can form a difference that cancels both the laser-frequency noise and the clock noise. Defining
\begin{align}
    \xi_{1}\equiv \varepsilon_1-\tau_1,
\end{align}
 we obtain
\begin{align}
    \xi_{1}^{\text{noise}}\simeq 4\pi\nu_{1^{\prime}}\left(\mathbf{n}_{21}\cdot\boldsymbol{\delta}_1-\mathbf{n}_{21}\cdot\boldsymbol{\Delta}_1\right)+N^{\varepsilon}-N^{\tau}.
\end{align}

The laser-frequency noise and the clock noise are absent in this
combination. Therefore, for an ULDM-induced differential acceleration, the
corresponding signal contribution is
\begin{align}
    \xi_{1}^{\text{signal}}\simeq 4\pi\nu_{1^{\prime}}~\mathbf{n}_{21}\cdot\left(\delta \boldsymbol{x}_{\text{TM}}-\delta \boldsymbol{x}_{\text{OB}}\right).
\end{align}
Below we derive the sensitivity of $\xi$ to ULDM. 
Throughout this section we consider a stationary detector, a monochromatic ULDM field, and average over the ULDM propagation direction $\boldsymbol{k}$.

We first consider the case for nonvanishing $d_g$, while setting $d_{\hat{m}}=d_e=0$ for simplicity. From Eq.~\eqref{eq:EOM}, the
composition-dependent part of the acceleration is proportional to
$d_g Q_{\hat{m}}\nabla\Phi$. Therefore, the relative TM--OB response is
controlled by the charge difference
\begin{align}
    \Delta Q_{\hat{m}}
    \equiv
    Q_{\hat{m}}^{\mathrm{OB}}
    -
    Q_{\hat{m}}^{\mathrm{TM}} .
\end{align}

After converting to relative-frequency units, the signal can be written as
\begin{align}
    \xi_1(t)=\frac{i\kappa d_g\Delta Q_{\hat{m}}}{m}\mathbf{n}_{21}\cdot\nabla\Phi,
\end{align}
For a monochromatic ULDM field with local density $\rho$, the field
amplitude satisfies $\Phi_0\simeq \sqrt{2\rho}/m$. If the typical ULDM
velocity dispersion is denoted by $\sigma$, then $|\boldsymbol{k}|\simeq m\sigma$. Direction averaging gives
\begin{align}
    \left\langle
    \left(
        \mathbf{n}_{21}\cdot\nabla\Phi
    \right)^2
    \right\rangle_{\hat{\boldsymbol{k}}}
    =
    \frac{1}{3}|\boldsymbol{k}|^2 \frac{\Phi_0^2}{2}
    \simeq
    \frac{\rho\sigma^2}{3}.
\end{align}
Here the extra factor $1/2$ comes from the fact that the physical ULDM
field $\Phi$ is real: the signal power is computed from the squared real
oscillating field, whose time average gives
$\langle \cos^2(.)\rangle=1/2$.
With the one-sided PSD convention
\begin{align}
    P_s(f)\equiv \frac{2|\tilde{\xi}(f)|^2}{T},
\end{align}
the signal PSD is therefore
\begin{align}
    P_s(f)
    =\begin{cases}\label{eq:xiPSDT}
        \dfrac{2\kappa^2 d_g^2 \Delta Q_{\hat{m}}^2 \sigma^2 \rho}{3m^2}\,T, & f=f_c,\\[6pt]
        0, & f\neq f_c,
    \end{cases}
\end{align}
where $T$ is the observation time, $\tilde{\xi}(f)$ denotes the Fourier transform of the data stream $\xi(t)$, and $f_c=m/2\pi$ is the Compton frequency. For comparison, we can calculate the corresponding $X$-channel signal PSD,
\begin{align}
    P_X\left(f\right)=&128\sin^2\left(2\pi fL\right)\sin^4\left(\pi fL\right) \nonumber \\
    &\times\frac{\kappa^2{(1-Q_{\hat{m}}^{\mathrm{TM}})^2} d_g^2 \sigma^2 \rho}{3m^2} T,\;f=f_c\label{eq:XPSD}
\end{align}
where $L$ is arm length of SWGDs.

We adopt the LISA design in which each TM is formed from $73\%$ gold and $27\%$ platinum. The OB composition is more involved: according to the components listed in Table  {1} of Ref.~\cite{racca_lisa_2010}, we model a $450~\mathrm{kg}$ spacecraft as containing $83~\mathrm{kg}$ of carbon, $104.3~\mathrm{kg}$ of copper, with the remaining mass taken to have the same composition as aluminum. Under these assumptions, the corresponding effective charges are
\begin{align}
    Q_{\hat{m}}^{\mathrm{TM}}\approx 0.0853,
    \qquad
    Q_{\hat{m}}^{\mathrm{OB}}\approx 0.0805.
\end{align}

The displacement noises associated with $\boldsymbol{\delta}$ and $\boldsymbol{\Delta}$, and OMS noise are characterized by the following (one-sided) amplitude spectral densities:
\begin{align}
    \sqrt{S_{\mathrm{acc}}}
    =&\,
    \frac{s_{\mathrm{acc}}}{2\pi f c}\left[\frac{1}{\sqrt{\mathrm{Hz}}}\right]
    \sqrt{1+\left(\frac{0.4\times10^{-3}}{f}\right)^2}
    \nonumber\\
    &\times
    \sqrt{1+\left(\frac{f}{8\times10^{-3}}\right)^4},\\
    \sqrt{S_{\Delta}}
    =&\,
    \frac{5.0\times10^{-15}}{2\pi f c}
    \left[\frac{1}{\sqrt{\mathrm{Hz}}}\right],
    \\
    \sqrt{S_{\mathrm{oms}}}
    =&\,
    s_{\mathrm{oms}}\frac{2\pi f}{ c}\left[\frac{1}{\sqrt{\mathrm{Hz}}}\right]
    \sqrt{1+\left(\frac{2\times10^{-3}}{f}\right)^2}.
\end{align}
Although $\xi$ does not explicitly contain the optical-path noise, we treat it as being of the same order of magnitude as the shot noise. Therefore, we adopt the analytical OMS noise PSD above to characterize the shot-noise level. Also, for illustration, we take $S_{\Delta}$ from the LISA Pathfinder measurement at $1\,\mathrm{mHz}$ \cite{lisa_pathfinder_collaboration_lisa_2019}. The instrumental parameters adopted for the experiments considered here are listed in Table~\ref{tab:detectorparameter}.
\begin{table}
    \centering
    \begin{tabular}{lccc}
         \toprule
         & LISA & Taiji  & BBO\\
         \hline
         $L$ ($10^9 \mathrm{m}$) & 2.5 & 3 & 0.05 \\
         $s_{\mathrm{acc}}$($10^{-15}\mathrm{m/s^2}$) & 3 & 3 & $3\times10^{-2}$\\
         $s_{\mathrm{oms}}$($10^{-12}\mathrm{m}$) & 15 & 8 & $1.4\times10^{-5}$\\
         \hline
    \end{tabular}
    \caption{Key instrumental parameters adopted for the three space-based gravitational-wave detector concepts considered in this work, including the arm length $L$, the acceleration-noise level parameter  $s_{\mathrm{acc}}$, and the optical metrology system noise parameters $s_{\mathrm{oms}}$ for LISA, Taiji, and BBO.}
    \label{tab:detectorparameter}
\end{table}
For BBO, the frequency-dependent factors in the square brackets are neglected~\cite{corbin_detecting_2006}.

For $\tau_c>T$, the signal-to-noise ratio (SNR) is defined as
\begin{align}
    \text{SNR}=\frac{P_s(f_c)}{P_n(f_c)}
\end{align}
where $P_n=S_{\mathrm{acc}}+S_{\mathrm{oms}}+S_{\Delta}$. While for $\tau_c<T$, ULDM can no longer be regarded as a coherent monochromatic field, and the PSD Eq.\eqref{eq:xiPSDT} approximation is modified to \cite{budker_proposal_2014}
\begin{align}
    P_s^{\text{mod}}(f_c)\simeq \frac{2\kappa^2 d_g^2 \Delta Q_{\hat{m}}^2 \sigma^2 \rho}{3m^2}\tau_c,
\end{align}
and the SNR is defined as
\begin{align}
    \text{SNR}=\frac{P_s(f_c)}{P_n(f_c)\sqrt{T/\tau_c}}.
\end{align}
We define the sensitivity curve as the minimum coupling strength that yields $\mathrm{SNR}=1$ for an observation time of one year. 

In the case of electromagnetic coupling $d_e$, we can follow similar steps. We can replace
\begin{align}
    d_g\Delta Q_{\hat{m}}
    \longrightarrow
    d_e\Delta Q_e ,
\end{align}
where
\begin{align}
    \Delta Q_e
    \equiv
    Q_e^{\mathrm{OB}}
    -
    Q_e^{\mathrm{TM}} .
\end{align}
The corresponding signal PSD is
\begin{align}
    P_s^{e}(f)
    =
    \begin{cases}
        \dfrac{
        2\kappa^2 d_e^2
        \Delta Q_e^2
        \sigma^2\rho
        }{
        3m^2
        }\,T,
        & f=f_c,\\[8pt]
        0,
        & f\neq f_c .
    \end{cases}
    \label{eq:xiPSDTe}
\end{align}
For the same TM and OB material model, the effective electromagnetic
charges are
\begin{align}
    Q_e^{\mathrm{TM}}
    \approx 0.0043,
    \qquad
    Q_e^{\mathrm{OB}}
    \approx 0.0016 .
\end{align}
Although $Q_e$ itself is smaller than $Q_{\hat{m}}$, the relevant quantity
for the $\xi$ channel is the charge difference between the OB and the TM.
Since $\Delta Q_e$ and $\Delta Q_{\hat{m}}$ are of comparable order, the
$\xi$-based sensitivities to $d_e$ and $d_g$ are expected to lie on a
similar scale, up to the corresponding noise level and transfer-function
factors.

\begin{figure}[t]
    \centering
    \includegraphics[width=1\linewidth]{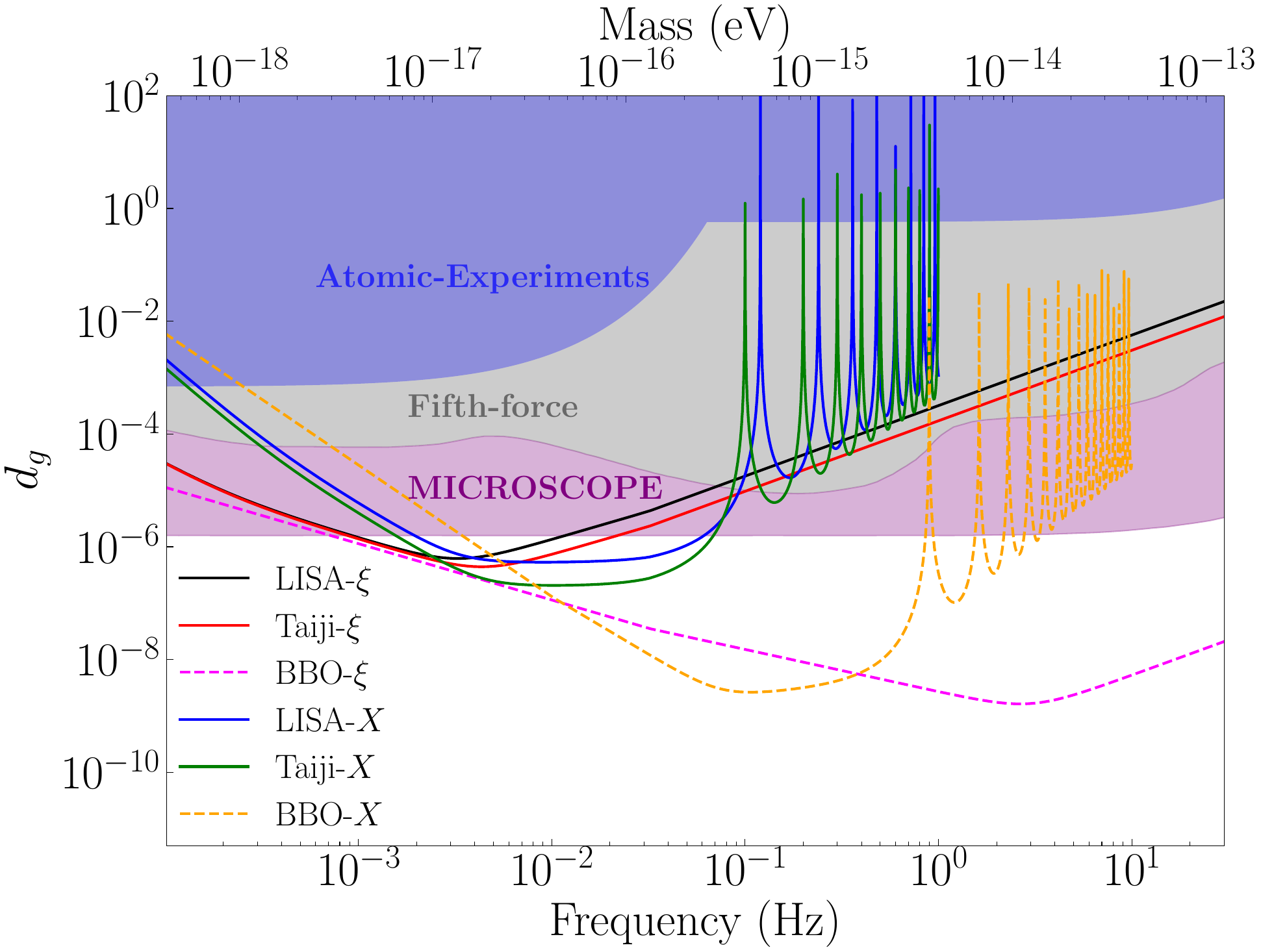}
    \caption{Sensitivity to the dilaton--gluon coupling $d_g$ as a function of frequency for LISA, Taiji, and BBO. The black, blue, and magenta curves are obtained from the $\xi$ observable, while the red, cyan, and orange curves show the corresponding sensitivities from the $X$-channel. The shaded regions indicate existing constraints from MICROSCOPE~\cite{PhysRevLett.119.231101,PhysRevLett.129.121102}, Fifth-force searches~\cite{fischbach1996yearsfifthforce,Lee_2020,PhysRevLett.124.051301}, and atomic experiments~\cite{dzuba_constraints_2024}.}
    \label{fig:sensitivity_dg}
\end{figure}

\begin{figure}[t]
    \centering
    \includegraphics[width=1\linewidth]{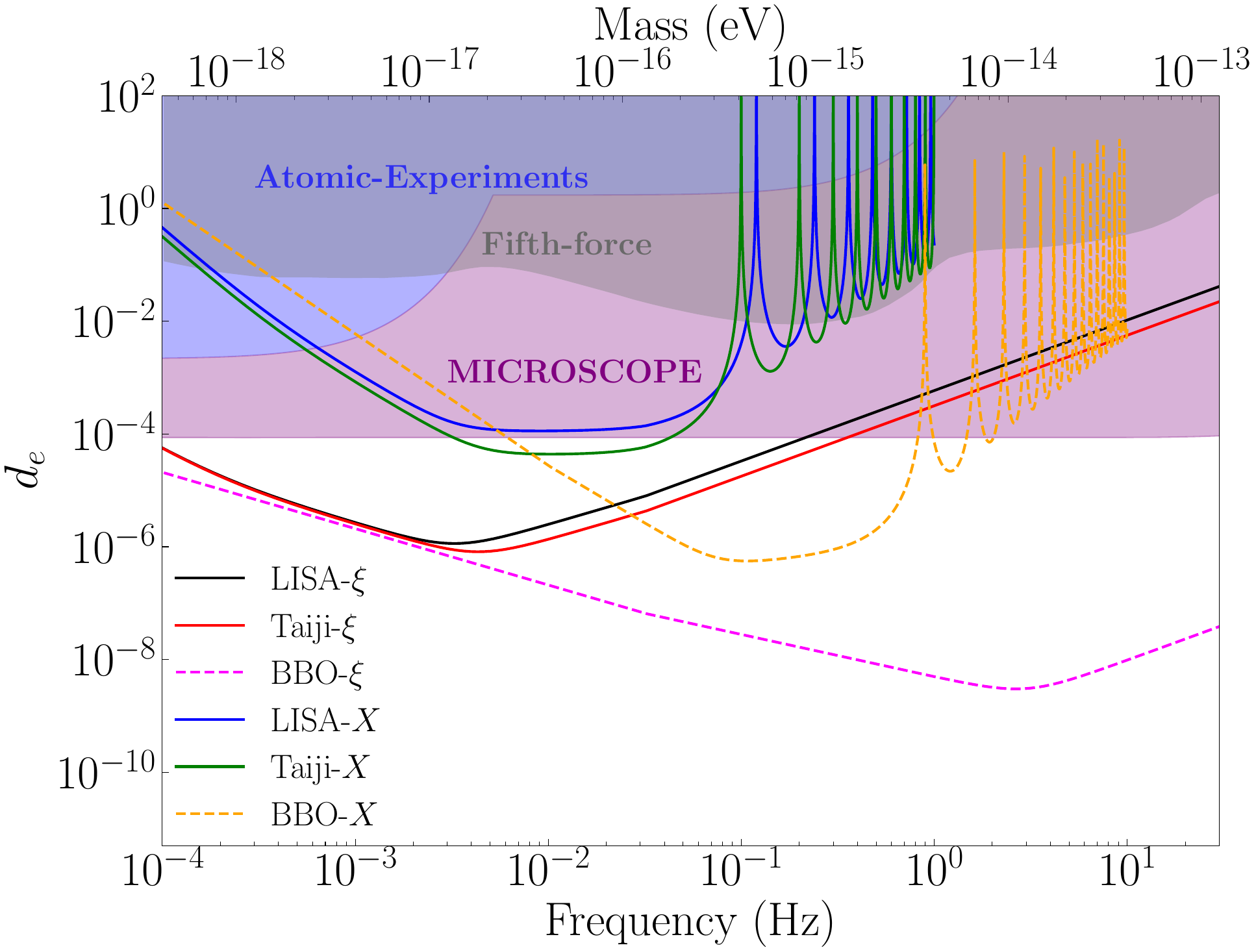}
    \caption{Sensitivity to the dilaton--electron coupling $d_e$. Here the shaded exclusion regions again show the current bounds from MICROSCOPE, Fifth-force searches, and atomic experiments, where the atomic bounds can likewise be interpreted as constraints on a fifth-force-like interaction.}
    \label{fig:sensitivity_de}
\end{figure}

In Figs.~\ref{fig:sensitivity_dg} and \ref{fig:sensitivity_de} we show the sensitivities to $d_g$ and $d_e$ for LISA, Taiji, and BBO. For each mission, we compare the $\xi$ observable with the standard Michelson $X$ channel, and we also overlay three classes of existing bounds: MICROSCOPE equivalence-principle tests~\cite{PhysRevLett.119.231101,PhysRevLett.129.121102}, macroscopic fifth-force searches~\cite{fischbach1996yearsfifthforce,Lee_2020,PhysRevLett.124.051301}, and atomic constraints~\cite{dzuba_constraints_2024}. Although the atomic bounds arise from precision atomic measurements, they are also interpreted through the effective fifth-force framework, so they are best viewed as a complementary fifth-force probe.

One aspect highlighted in Figs.~\ref{fig:sensitivity_dg} and \ref{fig:sensitivity_de} is the better performance of the $\xi$ observable in some frequency ranges. In the most sensitive band of LISA and Taiji, the $\xi$-curves are already competitive with the $X$-curves, while toward lower frequencies they become clearly better. The reason is that $\xi$ directly measures the local differential motion between the TM and the OB, therefore its response does not contain factors $\sin(2\pi fL)$ or $\sin(\pi fL)$ that appear in Eq.~\eqref{eq:XPSD}. For the X channel, the factor $\sin^2(2\pi fL)$ is canceled with its noise counterpart, whereas the factor $\sin^4(\pi fL)$ remains~\cite{yu_sensitivity_2023}. Consequently, the sensitivity of X channel at low-frequency inherits this residual suppression $\sin^4(\pi fL)$. 

We also notice that the relative improvement of $\xi$ over $X$ is much better for $d_e$ than $d_g$. This can be explained as follows. Eq.~\eqref{eq:amdQ} shows that the electromagnetic coupling $d_e$ enters the test-mass acceleration accompanied by the dilaton charge $Q_e$. Because $Q_e$ is far smaller than the gluonic prefactor $\sim 1$, the effect of $d_e$ in the $X$ channel is suppressed by roughly three orders of magnitude compared to that of $d_g$. In contrast, for the $\xi$ observable, both the $d_g$ and $d_e$ sensitivities are controlled by composition-dependent charge differences. Therefore, both channels are subject to similar charge suppression, and the relative performance of $\xi$ becomes more favorable in the measurement of $d_e$ than in the standard $X$ channel. 

The sensitivity of $\xi$ considered above originates from only one local differential
acceleration between the OB and the TM. In a triangular constellation, we can construct six $\xi$ observables. Considering that these six data streams have statistically independent noise and comparable response to the ULDM-induced
TM--OB differential motion, they can be combined for analysis. The total SNR will be increased by a factor $\sqrt{6}$, $
    \mathrm{SNR}_{\mathrm{total}} = \sqrt{6}\, \mathrm{SNR}_{\xi}.
$
Equivalently, the minimum detectable signal amplitude, and hence the
corresponding coupling sensitivity for a response linear in the coupling,
is improved by a factor of $\sqrt{6}$ compared with a single $\xi$ channel.

\section{Discussions}\label{sec:analysis}

 A number of studies have investigated the prospects for directly detecting ULDM through its imprint on the SGWD response. The key lesson of the present work is that observability is not determined solely by whether an ULDM-induced perturbation exists at the single-link level; it is determined by the structure of the single-link response and by whether that structure survives the standard calibration and TDI pipeline. To make this point explicit, we now summarize several representative classes of ULDM couplings to SGWDs and compare their single-link responses.

\textbf{Class  {1}: Laser-noise-like endpoint response.} If ULDM induces oscillations of fundamental constants that modulate instrumental parameters, such as the laser frequency discussed above, the corresponding single-link response is given by Eq.~\eqref{eq:etanu}. This is the class of signals that is structurally degenerate with laser phase noise.

\textbf{Class  {2}: Directional Doppler response.} If ULDM couples to the TMs and induces an effective force, the observable signal arises from the ULDM-driven velocity fluctuations of the TMs, which manifest as Doppler shifts in the received laser light~\cite{morisaki_detectability_2019,yu_sensitivity_2023}. In this scenario, the corresponding single-link response is described by Eq.~\eqref{eq:eta_nrs}.

\textbf{Class  {3}: Propagation-induced response.} If the relevant ULDM-SM interaction is an axion-like-particle (ALP) coupling to electromagnetism, the dominant effect may instead originate from birefringence or phase modulation accumulated during laser propagation through the ALP background field. In this case, the induced perturbation is not associated with test-mass motion but with the field-dependent modification of the optical path, leading to a distinct single-link response \cite{yao_prospects_2025,gue_probing_2025,liu_detectability_2026}
\begin{align}\label{eq:eta_ers}
    \eta_{rs}(t)=\epsilon_{rs} \xi \left[a(t)-\mathcal{D}_{rs}a(t)\right],
\end{align}
where $a(t)$ is ALP field, $\xi$ is a time-independent quantity and $\epsilon_{rs}$ is a three-dimensional antisymmetric second-order tensor represented as a $3\times 3$ matrix satisfying
\begin{align}
    \epsilon_{12}=\epsilon_{23}&=\epsilon_{31}=1,\nn\\
    \epsilon_{rs}&=-\epsilon_{sr}.
\end{align}

\textbf{Class  {4}: Gravitational redshift and metric response.} ULDM can also affect SGWDs by sourcing a weak, time-dependent gravitational field in the Solar neighborhood. The measured fractional-frequency shift is determined by the gravitational-potential difference between the reception and emission events and by the GW-like transverse-traceless metric perturbation projected along the link~\cite{yu_detecting_2024}. Then, to first order in the metric perturbations,
\begin{align}\label{eq:eta_GW}
    \eta_{rs}(t)&\simeq\left[\Phi^{\mathrm{G}}(t,\boldsymbol{x}_{r})-\mathcal{D}_{rs}\Phi^{\mathrm{G}}(t,\boldsymbol{x}_{s})\right]\nn\\
    &+\frac{1}{2}\hat{n}_{rs}^{a}\hat{n}_{rs}^{b}\left[h_{ab}^{\mathrm{TT}}(t,\boldsymbol{x}_{r})-\mathcal{D}_{rs}h_{ab}^{\mathrm{TT}}(t,\boldsymbol{x}_{s})\right]\nn\\
    &+\hat{n}_{rs}\cdot\left(\left.\frac{\dd \boldsymbol{x}}{\dd t}\right|_{t,\boldsymbol{x}_r}-\mathcal{D}_{rs}\left.\frac{\dd \boldsymbol{x}}{\dd t}\right|_{t,\boldsymbol{x}_s}\right).
\end{align}
Here $\Phi^{\mathrm{G}}$ is the weak gravitational potential perturbation sourced by ULDM, $h_{ab}^{\mathrm{TT}}$ is the GW-like transverse–traceless part of the metric perturbation (present if ULDM sources anisotropic stress), and ${\dd \boldsymbol{x}}/{\dd t}$ are the small gravity-induced velocities of the free-falling TMs. The first term is an endpoint gravitational redshift, the second is the GW-like strain projection along the arm, and the third is the ordinary Doppler shift from endpoint motion.

For signals of the form given in Eqs.~\eqref{eq:eta_nrs}, \eqref{eq:eta_ers}, and \eqref{eq:eta_GW}, the situation is markedly different from that of Eq.~\eqref{eq:etanu} after TDI. Even in the simplest equal-arm approximation, the corresponding TDI-processed signal is not strongly suppressed, whereas the result in Eq.~\eqref{eq:X1.5depressed} vanishes identically in that limit. As a consequence, physically meaningful information can still be extracted from the TDI observables.

The origin of this difference can be traced back to the structure of the single-link signal. For the signal induced by oscillations of fundamental constants considered in this work, the response is neither directional nor accumulated along the light path. As a result, the corresponding single-link contribution has the same operator form as the laser phase noise, namely $s_{rs}(t)=\mathcal{D}_{rs}p_s(t)-p_r(t)$. In contrast, Eq.~\ref{eq:eta_GW} and ULDM-induced signals associated with TM motion or laser propagation contain factors such as $\hat{n}_{rs}$ or $\epsilon_{rs}$ that are not acted upon by the delay operator $\mathcal{D}_{rs}$. These factors preserve directional information and prevent the signal from being rewritten in the laser-noise-like form. Such contributions therefore survive TDI, whereas a signal of the form in Eq.~\eqref{eq:etanu}, although present in the individual links, is inevitably suppressed or canceled together with the laser phase noise.

The above comparison provides a practical criterion: ULDM-induced contributions that are structurally degenerate with the laser-noise or clock-noise channels are removed by the corresponding elimination procedure, whereas contributions carrying directional or propagation information can survive into the final observables. For the laser-frequency-variation signal considered in this work, the relevant single-link response is structurally indistinguishable from laser phase noise and is therefore eliminated by TDI. This implies that SGWDs relying on TDI combinations, such as LISA and Taiji, are intrinsically insensitive to this class of signals. 
Others that do not rely on the same mechanism may still probe ULDM signals with this structure. This highlights the complementarity between different classes of precision interferometric experiments in the search for ULDM.

\section{Conclusion}\label{sec:conclusion}
In this work, we have investigated how an ULDM field, modeled as a coherent oscillation that induces periodic variations of fundamental constants, can imprint signatures on space-based laser interferometers, including LISA, Taiji and BBO. Starting from the single-link measurements, including both the science readouts and the auxiliary sideband readouts, we have traced how different ULDM-induced effects propagate through the standard data processing, including laser and clock noise elimination.

Our main conclusion is that the detectability of an ULDM-induced effect is controlled by the structure of its single-link response. We have identified that ULDM-induced signals that retain directional or propagation information, such as the responses in Eqs.~\eqref{eq:eta_nrs}, \eqref{eq:eta_ers}, and \eqref{eq:eta_GW}, are not structurally degenerate with the laser-noise channel and can therefore survive into the final observables. For the laser-frequency-variation signal, the induced contribution is neither directional nor accumulated along the propagation path, and it shares the same operator form as the laser phase noise. It is therefore strongly suppressed in the final interferometry channels. The same structural viewpoint also explains why ULDM-induced effects on clocks are removed, or strongly suppressed, by the standard sideband-based clock noise elimination. 

In contrast, we have illustrated that the $\xi$ observable directly tracks the differential motion between the TM and the OB, therefore providing a complementary probe of ULDM. Its sensitivity is comparable to, or in some physical cases even better than, the standard Michelson interferometry channel in the relevant frequency ranges. 


\begin{acknowledgements}
This work is partly supported by the National Key Research and Development Program of China (Grant No.~2021YFC2201901), 
the National Natural Science Foundation (Grant No.12547104),
and the Fundamental Research Funds for the Central Universities. 
\end{acknowledgements}

\appendix
\section{Time-delay interferometry and Michelson $X$ channel}
\label{appendix:X}

Laser phase noise is many orders of magnitude larger than both the gravitational-wave signal and the secondary instrumental noises. A direct Michelson-type subtraction of the raw one-way measurements would cancel this noise only if the two optical paths had exactly the same light-travel time. This condition is not satisfied for space-based detectors, whose arm lengths are unequal, and slowly time-dependent.

Time-delay interferometry (TDI)~\cite{tinto_time-delay_2020, armstrong_time-delay_1999, estabrook_time-delay_2000, Wang:2020pkk} and earlier attempts~\cite{10.1117/12.293329} solve this problem by synthesizing equal-path interferometers from delayed one-way measurements. The basic idea is to compare two optical paths that start and end at the same spacecraft and contain the same laser-noise samples at the same retarded times. If the effective optical paths are matched, the laser phase noise cancels algebraically.

In this appendix we explicitly use the standard terminology $X_{1.5}$ and $X_{2.0}$. $X_{1.5}$ is the unequal-arm Michelson combination that cancels laser phase noise exactly for time-independent arms. $X_{2.0}$ is a higher-order Michelson combination designed to cancel laser phase noise when the arm lengths are slowly time-dependent, by matching the two effective optical paths up to first order in the arm-length variation.

We define the delay operator by
\begin{align}
    \mathcal{D}_{ij}f(t)\equiv f\!\left(t-L_{ij}(t)\right),
\end{align}
where $L_{ij}(t)$ is the arm length from spacecraft $i$ to spacecraft $j$. For LISA/Taiji, $\dot L_{ij}\sim10^{-9}$.

For later use, we also note the physical origin of the residual terms. When the arm-lengths are constant, different delay operators commute. When they vary with time, the order of the delays matters. For two generic delays, one has schematically
\begin{align}
    \left[\mathcal{D}_{a},\mathcal{D}_{b}\right]f(t)
    =
    \mathcal{O}(\dot L)\,\dot f ,
\end{align}
which means that the mismatch between two optical paths is controlled by the arm-length variation rate. This is why a residual contribution can survive in $X_{1.5}$ at order $\dot L$, while $X_{2.0}$ is designed to cancel it through this order.

\subsection{$X_{1.5}$: unequal-arm Michelson combination}

\begin{figure*}[t]
  \centering
  \begin{subfigure}[t]{0.48\textwidth}
    \centering
    \includegraphics[width=\linewidth]{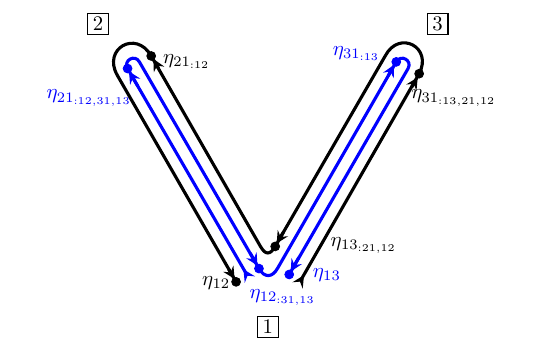}
    \caption{$X_{1.5}$ optical-path construction.}
    \label{fig:X1.5}
  \end{subfigure}
  \hfill
  \begin{subfigure}[t]{0.48\textwidth}
    \centering
    \includegraphics[width=\linewidth]{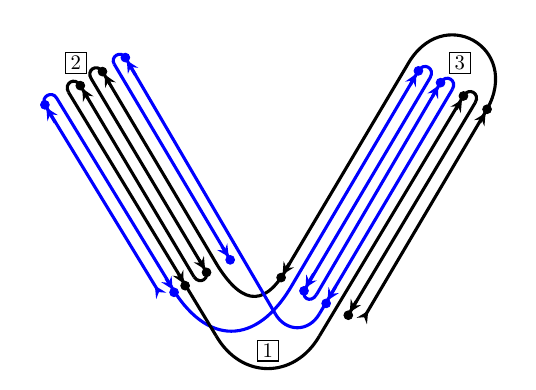}
    \caption{$X_{2.0}$ optical-path construction.}
    \label{fig:X2.0}
  \end{subfigure}

  \caption{
  Schematic illustration of optical paths in the X channel. Squares denote the three spacecrafts. Blue and black curves denote the two synthetic optical paths being compared. Each segment between adjacent dots corresponds to one delayed one-way measurement. We use the shorthand $\eta_{ab_{:cd}}\equiv \mathcal{D}_{cd}\eta_{ab}$.
  }
  \label{fig:X_channel}
\end{figure*}

$X_{1.5}$ is the four-link unequal-arm Michelson combination centered on spacecraft 1. It compares the two round-trip optical paths
\[
1\rightarrow 3\rightarrow 1\rightarrow 2\rightarrow 1
\qquad \text{and} \qquad
1\rightarrow 2\rightarrow 1\rightarrow 3\rightarrow 1,
\]
and is defined algebraically as~\cite{tinto_time-delay_2020} 
\begin{align}
   & X_{1.5}
    =
    \eta_{13}
    +\mathcal{D}_{13}\eta_{31}
    +\mathcal{D}_{13}\mathcal{D}_{31}\eta_{12}
    +\mathcal{D}_{13}\mathcal{D}_{31}\mathcal{D}_{12}\eta_{21}
    \nn\\
    &
    -\eta_{12}
    -\mathcal{D}_{12}\eta_{21}
    -\mathcal{D}_{12}\mathcal{D}_{21}\eta_{13}
    -\mathcal{D}_{12}\mathcal{D}_{21}\mathcal{D}_{13}\eta_{31}.
    \label{eq:X1.5}
\end{align}
This expression can be understood as the difference between two synthesized optical paths, see Fig.~\ref{fig:X_channel}. For static arms, the two paths have the same total light-travel time, and the laser phase noise is sampled at the same retarded times with opposite signs.

Using the one-way frequency-shift expression in Eq.~\eqref{eq:etanu}, the laser-noise terms cancel pairwise in Eq.~\eqref{eq:X1.5}. The remaining ULDM-induced contribution can be written as
\begin{align}
    X_{1.5}
    =
    -
    \left(
    \mathcal{D}_{13}\mathcal{D}_{31}\mathcal{D}_{12}\mathcal{D}_{21}
    -
    \mathcal{D}_{12}\mathcal{D}_{21}\mathcal{D}_{13}\mathcal{D}_{31}
    \right)
    \frac{\delta\nu_1}{\nu}.
    \label{eq:X15_dnu}
\end{align}
This equation shows that the residual signal is determined by the difference between the two four-link delay sequences. The two four-link delay sequences are rewritten as
\begin{align}
    \mathcal{D}_{12}\mathcal{D}_{21}\mathcal{D}_{13}\mathcal{D}_{31}f(t)
    &=
    f\!\left(t-\delta_{a}\right),
    \label{eq:d1231}
    \\
    \mathcal{D}_{13}\mathcal{D}_{31}\mathcal{D}_{12}\mathcal{D}_{21}f(t)
    &=
    f\!\left(t-\delta_{b}\right),
    \label{eq:d1321}
\end{align}
with
\begin{widetext}
\begin{align}
    \delta_{a}
    &\equiv
    L_4
    -
    \left[
    \dot L_{31}\left(L_{13}+L_{12}+L_{21}\right)
    +\dot L_{13}\left(L_{12}+L_{21}\right)
    +\dot L_{21}L_{12}
    \right],
    \label{eq:d1231L}
    \\
    \delta_{b}
    &\equiv
    L_4
    -
    \left[
    \dot L_{21}\left(L_{12}+L_{13}+L_{31}\right)
    +\dot L_{12}\left(L_{13}+L_{31}\right)
    +\dot L_{31}L_{13}
    \right]. 
    \label{eq:d1321L}
    \\
    L_4&\equiv L_{12}+L_{21}+L_{13}+L_{31}\simeq4L .
\end{align}
\end{widetext}
Their difference is
\begin{align}
    \delta_{a}-\delta_{b}
    =&
    \left(\dot L_{12}+\dot L_{21}\right)
    \left(L_{13}+L_{31}\right)\nn\\
    -&
    \left(\dot L_{13}+\dot L_{31}\right)
    \left(L_{12}+L_{21}\right).
    \label{eq:delta1231_delta1321}
\end{align}
Therefore, if the arms are time-independent, all $\dot L_{ij}$ vanish and $ \delta_{a}=\delta_{b}=L_4 $.


Using Eq.~\eqref{eq:nuam}, the ULDM-induced fractional frequency modulation can be written as
\begin{align}
    \frac{\delta\nu_1}{\nu}
    =
    2\kappa\left(d_e+d_{m_e}\right)\Phi(t).
\end{align}
Substituting this relation into Eq.~\eqref{eq:X15_dnu}, one obtains
\begin{align}
    X_{1.5}
    \simeq
    -2\kappa\left(d_e+d_{m_e}\right)
    \left[
    \Phi\!\left(t-\delta_{b}\right)
    -
    \Phi\!\left(t-\delta_{a}\right)
    \right].
    \label{eq:X1.5phidelta}
\end{align}
We can expand at $t-L_4$ and keep terms linear in $\dot L_{ij}$,
\begin{align}
    \Phi\!\left(t-\delta_{b}\right)
    -
    \Phi\!\left(t-\delta_{a}\right)
    \simeq
    \left(\delta_{a}-\delta_{b}\right)
    \dot{\Phi}(t-L_4).
\end{align}
Substituting Eq.~\eqref{eq:delta1231_delta1321}, we finally obtain
\begin{widetext}
\begin{align}
    X_{1.5}
    &\simeq
    -2\kappa\left(d_e+d_{m_e}\right)
    \left[
    \left(\dot L_{12}+\dot L_{21}\right)
    \left(L_{13}+L_{31}\right)
    -
    \left(\dot L_{13}+\dot L_{31}\right)
    \left(L_{12}+L_{21}\right)
    \right]
    \dot{\Phi}(t-L_4).
    \label{eq:X1.5depressed}
\end{align}
\end{widetext}
This expression shows the key feature of $X_{1.5}$ for the ULDM-induced signal. If the arms are time-independent, then all $\dot L_{ij}$ vanish and the two effective four-link delays are identical. Hence the ULDM-induced contribution cancels exactly. For time-dependent arms, the two delay sequences no longer commute, and the residual signal is proportional to $\dot L_{ij}$. Therefore in $X_{1.5}$ the ULDM signal encoded in laser frequency is further suppressed by the small parameter $\epsilon_L\sim10^{-9}$ for LISA/Taiji.

\subsection{$X_{2.0}$: higher-order unequal-arm Michelson combination}

$X_{2.0}$ is constructed by further matching the two synthetic Michelson paths when the arm lengths vary slowly in time. Compared with $X_{1.5}$, it contains additional delayed one-way measurements such that the two effective optical paths agree not only for static arms, but also through first order in $\dot L_{ij}$, and is defined algebraically as~\cite{tinto_time-delay_2020}
\begin{widetext}
\begin{align}
    X_{2.0}
    &=
    \eta_{13}
    +\mathcal{D}_{13}\eta_{31}
    +\mathcal{D}_{13}\mathcal{D}_{31}\eta_{12}
    +\mathcal{D}_{13}\mathcal{D}_{31}\mathcal{D}_{12}\eta_{21}
    +\mathcal{D}_{13}\mathcal{D}_{31}\mathcal{D}_{12}\mathcal{D}_{21}\eta_{12}
    \nn\\
    &\quad
    +\mathcal{D}_{13}\mathcal{D}_{31}\mathcal{D}_{12}\mathcal{D}_{21}\mathcal{D}_{12}\eta_{21}
    +\mathcal{D}_{13}\mathcal{D}_{31}\mathcal{D}_{12}\mathcal{D}_{21}\mathcal{D}_{12}\mathcal{D}_{21}\eta_{13}
    +\mathcal{D}_{13}\mathcal{D}_{31}\mathcal{D}_{12}\mathcal{D}_{21}
    \mathcal{D}_{12}\mathcal{D}_{21}\mathcal{D}_{13}\eta_{31}
    \nn\\
    &\quad
    -\eta_{12}
    -\mathcal{D}_{12}\eta_{21}
    -\mathcal{D}_{12}\mathcal{D}_{21}\eta_{13}
    -\mathcal{D}_{12}\mathcal{D}_{21}\mathcal{D}_{13}\eta_{31}
    -\mathcal{D}_{12}\mathcal{D}_{21}\mathcal{D}_{13}\mathcal{D}_{31}\eta_{13}
    \nn\\
    &\quad
    -\mathcal{D}_{12}\mathcal{D}_{21}\mathcal{D}_{13}\mathcal{D}_{31}\mathcal{D}_{13}\eta_{31}
    -\mathcal{D}_{12}\mathcal{D}_{21}\mathcal{D}_{13}\mathcal{D}_{31}
    \mathcal{D}_{13}\mathcal{D}_{31}\eta_{12}
    -\mathcal{D}_{12}\mathcal{D}_{21}\mathcal{D}_{13}\mathcal{D}_{31}
    \mathcal{D}_{13}\mathcal{D}_{31}\mathcal{D}_{12}\eta_{21}.
    \label{eq:X2.0}
\end{align}
\end{widetext}
This longer combination can be viewed as follows: each four-link path is followed by the opposite four-link path, so that the mismatch caused by the noncommutativity of time delays is canceled to first order in $\dot L_{ij}$.

After the same cancellation of laser-noise terms and common scalar-induced terms, the residual ULDM contribution can be written as the difference between two eight-link delayed scalar-field samples,
\begin{align}
    X_{2.0}
    =
    -2\kappa\left(d_e+d_{m_e}\right)
    \left[
    \Phi\!\left(t-\delta_{c}\right)
    -
    \Phi\!\left(t-\delta_{d}\right)
    \right].
    \label{eq:X20_phi_delta}
\end{align}
The two eight-link delays are defined by
\begin{align}
    \mathcal{D}_{13}\mathcal{D}_{31}\mathcal{D}_{12}\mathcal{D}_{21}
    \mathcal{D}_{12}\mathcal{D}_{21}\mathcal{D}_{13}\mathcal{D}_{31}f(t)
    &=
    f\!\left(t-\delta_{c}\right),
    \label{eq:d13211231}
    \\
    \mathcal{D}_{12}\mathcal{D}_{21}\mathcal{D}_{13}\mathcal{D}_{31}
    \mathcal{D}_{13}\mathcal{D}_{31}\mathcal{D}_{12}\mathcal{D}_{21}f(t)
    &=
    f\!\left(t-\delta_{d}\right).
    \label{eq:d12311321}
\end{align}
Equivalently, using the four-link delays defined above,
\begin{align}
    \delta_{c}
    &=
    \delta_{a}(t-\delta_{b})+\delta_{b}(t),
    \label{eq:d13211231_def}
    \\
    \delta_{d}
    &=
    \delta_{b}(t-\delta_{a})+\delta_{a}(t).
    \label{eq:d12311321_def}
\end{align}
Expanding these expressions to first order in $\dot L_{ij}$ gives
\begin{align}
    \delta_{c}
    &=
    \delta_{a}
    +
    \delta_{b}
    -
    \dot L_4 L_4
    +
    \mathcal{O}(\dot L^2),
    \label{eq:d13211231L}
    \\
    \delta_{d}
    &=
    \delta_{b}
    +
    \delta_{a}
    -
    \dot L_4 L_4
    +
    \mathcal{O}(\dot L^2),
    \label{eq:d12311321L}
\end{align}
where
\begin{align}
    \dot L_4
    \equiv
    \dot L_{12}+\dot L_{21}+\dot L_{13}+\dot L_{31}.
\end{align}
Thus the two eight-link effective delays are identical up to first order,
\begin{align}
    \delta_{c}
    -
    \delta_{d}
    =
    \mathcal{O}(\dot L^2).
\end{align}
Substituting this result into Eq.~\eqref{eq:X20_phi_delta}, one obtains
\begin{align}
    X_{2.0}
    =
    \mathcal{O}(\dot L^2).
    \label{eq:X2.0depressed}
\end{align}
One can see that the ULDM-induced contribution that survives in $X_{1.5}$ at order $\dot L$ is further canceled in $X_{2.0}$. Therefore, after neglecting terms of order $\dot L^2$ and higher, $X_{2.0}$ contains no physically extractable ULDM signal from this contribution.

\bibliographystyle{apsrev4-2}
\bibliography{ref}

\end{document}